\documentclass[journal]{IEEEtran}
\usepackage[T1]{fontenc}
\usepackage{amsmath,amssymb,bm}
\usepackage{graphicx,booktabs,siunitx,xcolor}
\usepackage{cite}
\usepackage[colorlinks=true,linkcolor=blue!50!black,citecolor=blue!50!black,urlcolor=blue!50!black]{hyperref}
\graphicspath{{figures/}}
\input{labels_supplement_SI}
\newcommand{\citep}[1]{\cite{#1}}
\newcommand{\E}{\mathbf{E}}

\newcommand{\V}{\mathbf{V}}
\newcommand{\Vh}{\hat{\mathbf{V}}}
\newcommand{\bS}{\bm{\Sigma}}
\newcommand{\ValDev}{5.2}
\newcommand{\CovAexact}{0.92}
\newcommand{\CovAexactSixtyEight}{0.68}
\newcommand{\CovCurSmallTrueNF}{0.85}

\newcommand{\CovCurOnePct}{0.13}
\newcommand{\CovCurFivePct}{0.74}
\newcommand{\CovEHmeas}{0.92}
\newcommand{\CovEHrange}{0.94--0.97}

\newcommand{\CovEHwide}{0.92}
\newcommand{\AxialEH}{0.96}
\newcommand{\AxialE}{0.89}
\newcommand{\PredEmpEH}{1.01 (5--95\%: 0.98--1.04)}

\newcommand{\DepOurs}{1.054, 1.330, 1.382, 1.371}

\newcommand{\PredEmpE}{0.91 (5--95\%: 0.21--1.03)}

\newcommand{\HitShareGAGGxy}{8\%}
\newcommand{\HitShareGAGGz}{22\%}

\newcommand{\NTriplesExp}{4\,600\,098}

\newcommand{\NPairsExp}{15\,449\,794}
\newcommand{\NUniqueLORs}{506\,143}
\newcommand{\NSimTriples}{4119}
\newcommand{\NSimTrue}{912}

\newcommand{\RealParagraph}{\emph{Two-photon data.} Randoms make up 8.3\% of the prompt pairs. The fitted rod centers lie within 0.5~mm of the nominal positions; the axial centers are 3.5, 3.5, 4.0 and 3.5~mm and the lengths 4, 4, 5 and 4~mm for the 2, 3, 4 and 5~mm rods (1~mm grid). Relative to the 5~mm rod, the two-photon concentrations of the 2, 3 and 4~mm rods are 0.95, 0.92, 0.80 (0.94--0.95, 0.92 and 0.75--0.83 over the tube width, the rod length $\pm1$~mm and the axial center $\pm0.5$~mm). For nominal diameters and a uniform detector they would be 1; they absorb deviations of the actual rod volumes and of the detector normalization, which cancel in $\eta$ only to the extent that they affect both photon counts alike. The 0.80 of the 4~mm rod mostly reflects its longer fitted length (5 vs.\ 4~mm); with the nominal geometry it is 0.99, and $\eta$ changes little because the three-photon concentration moves with it.

\emph{Enhancement ratio.} With the corrected pixel map, the ROI pipeline gives $\eta=2.07, 1.08, 1.27$ for ROIs padded by half the FWHM and 2.82, 1.38, 1.31 for a full FWHM (Table~\ref{tab:real}), so the values depend strongly on the ROI size. The region model (main analysis: $E_\text{tot}\ge985$~keV, $t_{13}<40$~ns, calibrated covariance, 73\,254 events, of which 6674 are assigned to the rods, 12\,494 to the diffuse three-photon class, 26\,176 to scatter and 27\,914 to randoms) gives $\eta=$ 0.86 [0.75, 0.98] (2~mm), 0.68 [0.62, 0.76] (3~mm) and 0.98 [0.93, 1.05] (4~mm). Across the analysis variants with $E_\text{tot}\ge985$~keV or $\ge1000$~keV the medians range over 0.81--1.15, 0.63--0.79 and 0.96--1.05; using all energy sums, where scatter and randoms are 6 times more numerous, moves them to 0.55, 0.53, 0.85. The 3~mm rod is below 1 in every variant. The model-based values are therefore much closer to 1 than the ROI values for the smallest rod, but the 95\% intervals are statistical only: the spread across reasonable background and energy models is of the same size as the interval width, and the variants disagree most where the signal is weakest. Fig.~\ref{fig:real}g shows a second limitation: the share of 100--450~keV photons that a module records, relative to its share of 511~keV photons, varies between 0.32 and 1.42 across modules, so the three-photon and two-photon sensitivities of a rod are not proportional and do not cancel exactly in $\eta$. A first-order correction from these hit rates changes $\eta$ by $-14$\%, $-1$\%, $-5$\% for the 2, 3 and 4~mm rods; a normalization measurement is needed to replace it.}
\newcommand{\RealCaption}{Measured four-rod phantom (corrected pixel map). (a) Two-photon list-mode MLEM image of the prompt pairs. (b) Kernel deposition ($\bS_E+\bS_H$) of all prompt triples in the analysis box with $E_\text{tot}\ge985$~keV and $|z-3.5|\le8$~mm. (c) The same events weighted by their posterior probability of belonging to a rod (main region model). (d) Three-photon image of the ROI pipeline (energy-only kernels, all $t_{13}$). Square-root scale; dashed circles: fitted rod positions and nominal diameters. (e) Enhancement ratio of the 2, 3 and 4~mm rods by method (ROI: the pipeline of (d); model: the region model; dashed: nominal equal-ratio reference); Bayesian estimates with 95\% intervals. (f) Geant4: three-photon concentration ratios (truth 1) from the labeled true events, from the region model on all events, and from the ROI estimator. (g) Ratio of each module's share of photons in prompt triples to its share in pairs, a proxy for the response at 100--450~keV relative to 511~keV; labels mark the module that each rod faces.}

\newcommand{\GLGaussCov}{0.70}
\newcommand{\GLGaussCovSixtyEight}{0.44}
\newcommand{\GLMixCov}{0.94}
\newcommand{\GLMixCovSixtyEight}{0.67}

\newcommand{\GLMixPi}{0.68, 0.27, 0.05}
\newcommand{\GLMixC}{1.2, 5.4, 35.4}

\newcommand{\GLLatePrompt}{0.88}

\newcommand{\GLErrMed}{3.4}
\newcommand{\GLErrNinety}{9.0}
\newcommand{\GLErrNinetyNine}{20}
\newcommand{\GLNThreeStudy}{9\,000}
\newcommand{\GLLatePromptSimMeas}{0.88}

\newcommand{\GLRandPerTrueMeas}{4.1}

\newcommand{\GLRandPerTrueHigh}{8.2}

\newcommand{\RealRandPerTrue}{5.9}

\newcommand{\GLDiagLowNoScatTwo}{+23}
\newcommand{\GLDiagRefNoRandTwo}{+13}
\newcommand{\GLDiagRefNoScatTwo}{+7}
\newcommand{\GLDiagModNoRand}{+9, +12, +8}

\newcommand{\GLDiffMeasDLL}{5\,300--7\,800}

\newcommand{\GLDiffMeasTwo}{0.87}
\newcommand{\GLDiffMeasBaseTwo}{1.37}

\newcommand{\GLReferenceBBias}{+8.8, +0.7, $-$1.0}
\newcommand{\GLReferenceBCov}{0.53, 0.93, 0.90}

\newcommand{\GLReferenceGBias}{$-$8.3, $-$5.6, $-$3.0}

\newcommand{\GLReferenceMhBias}{+21.4, $-$4.9, $-$7.5}
\newcommand{\GLReferenceMhCov}{0.00, 0.83, 0.30}

\newcommand{\GLReferenceMfBias}{+48.1, +6.6, $-$1.5}
\newcommand{\GLReferenceMfCov}{0.00, 0.60, 1.00}

\newcommand{\GLLowCountsBBias}{+28.2, +6.9, +1.4}

\newcommand{\GLNoBackgroundBBias}{+1.9, +1.4, $-$1.0}
\newcommand{\GLNoBackgroundBCov}{0.90, 0.97, 0.93}

\newcommand{\GLModuleResponseBBias}{+24.5, +19.4, +12.5}

\newcommand{\GLLargeTemplatesBCov}{0.80, 0.90, 0.87}

\newcommand{\GLLargeTemplatesBTwo}{+7.3}

\newcommand{\GLRandomsMeasuredMhTwo}{1.92}

\newcommand{\GLRandomsMeasuredMfTwo}{2.55}

\newcommand{\GLRandomsMeasuredGTwo}{0.84}

\newcommand{\GLRandomsMeasuredBTwo}{1.41}

\newcommand{\GLRandomsMeasuredNTwo}{1.05}
\newcommand{\GLRandomsMeasuredNThree}{0.93}

\newcommand{\GLRandomsHighMhTwo}{2.32}

\newcommand{\GLRandomsHighMfTwo}{3.05}

\newcommand{\RealMix}{1.02 [0.92, 1.14], 0.74 [0.67, 0.80], 1.02 [0.96, 1.08]}

\newcommand{\RealMixDiffTwo}{0.84}

\newcommand{\RealMixDm}{1.07 [0.95, 1.19], 0.76 [0.69, 0.83], 1.03 [0.97, 1.09]}

\newcommand{\RealMainDmTwo}{0.92}

\newcommand{\RealDiffuseDeltaLL}{3\,177}

\newcommand{\RealMOneDm}{2.28 and 3.00}
\newcommand{\RealSameModAll}{73.6}
\newcommand{\RealSameModBox}{10.1}

\newcommand{\RealLateFlat}{5}

\providecommand{\GLMLNSettings}{10}
\providecommand{\GLMLMaxBiasDiff}{0.1}
\providecommand{\GLMLRmseRatio}{0.99--1.02}
\providecommand{\GLMLReferenceBias}{$+8.9$, $+0.7$, $-1.0$}
\providecommand{\GLMLReferenceCov}{0.50, 0.93, 0.93}

\providecommand{\RealMLMaxDiff}{0.008}
\providecommand{\RealMLMix}{1.024, 0.737, 1.021}
\providecommand{\RealMLMixTwoCI}{[0.91, 1.15]}
\providecommand{\RealBMixTwoCI}{[0.92, 1.14]}
\providecommand{\RealNEventsSel}{73\,254}
\providecommand{\RealNGroupsSel}{69\,205}
\providecommand{\RealClusterPostRange}{0.91--1.08}

\providecommand{\CZLow}{2.32}
\providecommand{\CZHigh}{3.97}

\providecommand{\FRRefBRmse}{0.104, 0.035, 0.027}
\providecommand{\FRRefGSRmse}{0.068, 0.026, 0.019}

\providecommand{\FRNoBgBRmse}{0.054, 0.036, 0.026}

\providecommand{\FRNoBgHSRmse}{0.027, 0.019, 0.015}
\providecommand{\FRNoBgPHSRmse}{0.024, 0.017, 0.015}

\providecommand{\FRStrBRmse}{0.130, 0.032, 0.030}
\providecommand{\FRStrGSRmse}{0.124, 0.033, 0.030}

\providecommand{\FRStrPHSRmse}{0.156, 0.069, 0.028}

\providecommand{\RealPriorMaxShift}{0.03}

\providecommand{\PrBTwoRmseThou}{0.133}
\providecommand{\PrBTwoHSRmseThou}{0.133}
\providecommand{\PrBTwoGMRFRmseThou}{0.136}
\providecommand{\PrMLTwoRmseThou}{0.132}
\providecommand{\PrMLEMRmseThou}{0.132}
\providecommand{\PrGMRFvsGammaKLThree}{0.57}
\providecommand{\PrGMRFvsGammaKLThreeThou}{0.62}
\providecommand{\PrBTwoGMRFKLThree}{147}

\providecommand{\PrMLEMKLThree}{171}

\providecommand{\ChkNTwoMin}{10}

\providecommand{\PRRefBBias}{+8.8, +1.3, +0.5}

\providecommand{\PRRefBCov}{0.52, 0.88, 0.88}

\providecommand{\PRLowBTwo}{20.1}
\providecommand{\PRLowBTwoRange}{10.1--28.2}
\providecommand{\PRLowBTwoBsd}{6.6}

\providecommand{\PRRandBCov}{0.00, 0.44, 0.69}
\providecommand{\PRRandBTwo}{38.2}
\providecommand{\PRRandBTwoRange}{33.3--41.1}

\providecommand{\PRNoBgBBias}{+3.2, +1.2, +0.2}

\providecommand{\PRNoBgBCov}{0.88, 0.91, 0.93}

\providecommand{\PRMaxBiasDiff}{0.14}

\begin{document}
\title{Bayesian Image Reconstruction and Uncertainty Quantification for Three-Photon Positronium Ratio Imaging: Toward Radiotheranostic Applications}

\author{Hsin-Hsiung Huang,~\IEEEmembership{Senior Member,~IEEE,} Macoto Fujimoto, Berkin Uluutku, Andres F. Torres, Mizuki Uenomachi, and Kenji Shimazoe,~\IEEEmembership{Member,~IEEE}
\thanks{This work did not involve human subjects or animals in its research.}
\thanks{H.-H. Huang, B. Uluutku and A. F. Torres are with the Dr.\ Yvette Kanouff School of Data, Mathematical, and Statistical Sciences, University of Central Florida, Orlando, FL 32816, USA (e-mail: Hsin-Hsiung.Huang@ucf.edu).}
\thanks{M. Fujimoto is with the Department of Bioengineering, The University of Tokyo, Tokyo, Japan.}
\thanks{M. Uenomachi is with the Institute of Integrated Research, Institute of Science Tokyo, Tokyo, Japan.}
\thanks{K. Shimazoe is with the Department of Bioengineering and the Department of Nuclear Engineering and Management, The University of Tokyo, Tokyo, Japan.}
\thanks{Corresponding author: Hsin-Hsiung Huang.}}

\markboth{IEEE Transactions on Radiation and Plasma Medical Sciences}{Huang \MakeLowercase{\textit{et al.}}: Bayesian Reconstruction for Three-Photon Positronium Ratio Imaging}

\maketitle
\bstctlcite{IEEEexample:BSTcontrol}

\begin{abstract}
Three-photon annihilation imaging can estimate positronium decay ratios with pure positron emitters, including the imaging agents of radiotheranostic pairs, but quantitative reconstruction is limited by heterogeneous localization errors, selection-dependent sensitivity, and background coincidences. We develop a Bayesian reconstruction combining a three-dimensional event-response model with a Poisson likelihood and latent signal-background allocations. Photon-energy and interaction-position uncertainties enter the event covariance; a simulation-calibrated scale mixture accommodates heavy-tailed errors. Posterior sampling gives credible intervals for the three-to-two-photon enhancement ratio. We compare kernel deposition, voxel maximum-likelihood expectation-maximization, and matched regional maximum likelihood. At 3000 selected events in full-absorption simulation, the regional Bayesian estimator had ratio RMSE 0.067, versus 0.075 for kernel-deposition region-of-interest analysis. In a transport-based four-rod simulation without background, regional biases averaged at most 3.2\% over six independent simulations, whereas background and detector mismatch degraded performance. In a measured uniform phantom, the heavy-tailed regional model gave 1.02, 0.74, 1.02 for the 2, 3, 4~mm rods (5~mm reference; nominal 1). Matched regional maximum likelihood, the flat-prior posterior mode, agreed with the Bayesian estimates in every simulated setting (biases within 0.15 percentage points) and on the measured data (within 0.008), so gains derive from the regional likelihood. Under Poisson Kullback--Leibler loss, Gamma posterior means dominate maximum likelihood, appreciably only at small counts; horseshoe-type and smoothing priors changed voxel images but not regional ratios, whereas shrinking ratios across regions reduced error when regions shared a ratio. Posterior intervals quantify the uncertainty that treatment decisions require; calibration and background modeling remain limiting.
\end{abstract}


\section{Introduction}
\IEEEPARstart{P}{ositronium} imaging seeks information about the tissue microenvironment beyond radiotracer uptake. The term three-gamma imaging includes physically distinct measurements. Tashima and Yamaya \cite{tashima2024review} review methods that detect two annihilation photons together with a nuclear prompt gamma, including isotope discrimination, direct localization \cite{grignon2007,yoshida2020wgi}, and positronium lifetime imaging (PLI) \cite{moskal2021sciadv,qi2022}, which has been demonstrated in the human brain in vivo \cite{moskal2024sciadv}. In contrast, the present work reconstructs the three photons emitted in a single ortho-positronium annihilation. Their energies sum to approximately 1022~keV, and their directions satisfy momentum conservation. This process supports positronium ratio imaging (PRI) with pure positron emitters such as $^{18}$F \cite{fujimoto2026commphys,kacperski2004tmi}.

Ratio imaging is relevant to radiotheranostics: unlike lifetime imaging, which needs a prompt-gamma emitter such as $^{44}$Sc \cite{moskal2021sciadv}, it works with the positron-emitting agents that select patients for radioligand therapy, such as $^{18}$F- or $^{68}$Ga-labeled PSMA ligands before $^{177}$Lu-PSMA-617 \cite{sartor2021vision}, and $^{64}$Cu paired with $^{67}$Cu (Supplementary Section~\ref{S-sec:s-thera}). Treatment planning acts on thresholds, so lesion-level values need calibrated intervals, which we provide and test. Positronium is sensitive to oxygen \cite{shibuya2020oxygen}, the three-to-two-photon ratio was introduced as a PET diagnostic indicator \cite{jasinska2017} and has been proposed as a hypoxia biomarker \cite{moskal2026bams}, and $\beta$- but hardly $\alpha$-particle cell killing depends on oxygenation \cite{barendsen1966,poty2018alpha}, so a calibrated environment map could help choose between $\beta$ and $\alpha$ emitters; this is motivation, as the link between the three-photon ratio and tissue oxygenation remains to be established.

Our earlier PLI reconstructions model time-of-flight resolution in the likelihood \cite{huang2025pli,chen2024frontiers,uluutku2026conjugate,uluutku2026voxel}; here the same principle is applied to energy-dependent localization to estimate the three-to-two-photon activity ratio.

Fujimoto \emph{et al.} localized three-photon events directly from photon energies and interaction positions and estimated regional ratios \cite{fujimoto2026commphys}, building on earlier feasibility studies \cite{kacperski2004tmi,kacperski2005pmb}; the J-PET collaboration reconstructed three-photon images of an extended cylindrical phantom by positronium annihilation tomography \cite{moskal2021ncomms}; a companion study examines the point-spread function (PSF) of analytic localization \cite{fujimoto2026bams}. Other groups localize the vertex from times of flight by trilateration \cite{chug2026cpt}, from energies in closed form \cite{raczynski2026energy}, or from both with a QED prior in an event-by-event Bayesian estimate tested on simulated point sources \cite{raczynski2026trio}. This motivates a reconstruction of the ratio itself, with credible intervals, that models event-specific uncertainty, sensitivity, and background instead of treating events as exact sources.

Maximum likelihood estimation (MLE) is a criterion, maximum-likelihood expectation-maximization (MLEM) an algorithm \cite{shepp1982,parra1998}, and kernel deposition with ROI integration a different estimator; Bayesian emission reconstruction and latent-origin sampling are established \cite{green1990,higdon1997,sitek2008}. Our contribution is their formulation and evaluation for energy-localized three-photon PRI, including position-error propagation, selection-aware response, heavy-tailed event errors, and posterior propagation to a regional enhancement ratio.

We develop voxel and region models and evaluate them using full-absorption simulation, transport-based simulation, and a measured four-rod phantom. Comparison with the published-style ROI pipeline, voxel MLEM, and a matched regional MLE separates the effects of the response model from those of the prior; we also evaluate Poisson-adapted losses and shrinkage priors against matched ML.

\section{Materials and Methods}

\subsection{Vertex Estimator and Event Covariance}\label{sec:cov}
For hits $A,B,C$ and energies $\E$, momentum conservation gives the opening angle $\omega_{ij}$ between photons $i$ and $j$, $\cos\omega_{ij}=(E_k^2-E_i^2-E_j^2)/(2E_iE_j)$.
The vertex is the point that subtends these angles, the three-point resection problem, whose closed form is Tienstra's formula \citep{pierlot2014}
\begin{equation}
\V=\frac{k_AA+k_BB+k_CC}{k_A+k_B+k_C},\quad k_A=\big(\cot a-\cot\omega_{BC}\big)^{-1},
\label{eq:tienstra}
\end{equation}
with $a,b,c$ the angles of the hit triangle (Supplementary Section~\ref{S-sec:s-tienstra}). It equals the two-arc construction and places $\V$ in the plane of the hits.
Linearizing in the twelve inputs gives
\begin{equation}
\bS=\underbrace{J_E D_\sigma J_E^\top}_{\bS_E}+\underbrace{\textstyle\sum_{k}J_{P_k}\bS^\text{hit}_kJ_{P_k}^\top}_{\bS_H},
\label{eq:cov}
\end{equation}
where $D_\sigma=\mathrm{diag}(\sigma_A^2,\sigma_B^2,\sigma_C^2)$, $\sigma_k=E_k\,(\mathrm{FWHM}/E)(E_k)/2.355$ and $\bS^\text{hit}_k$ is the covariance of the first-interaction point about the pixel center (transverse variance $\approx2.5^2/12$~mm$^2$, energy-dependent depth variance from the simulation).
$\bS_E$ has rank two and lies in the hit plane; it is the covariance in current use.
$\bS_H$ adds the tilt of the decay plane and the axial error.
Exactly, $J_E\E=0$ (a common gain error cancels) and $J_{P_k}(P_k-\V)=0$ (depth along the photon direction does not propagate) (Supplementary Section~\ref{S-sec:s-jac}).

\subsection{Bayesian Image Reconstruction}\label{sec:bayes}
Let $y_i$ contain the measured photon energies, detector pixels, and reconstructed vertex of selected event $i$. For voxel $j$, let $q_j$ be its expected selected-event yield per unit activity, including acquisition exposure and all analysis cuts. Let $h(y\mid\mathbf x)$ be the density of the observation of a selected event from a decay at $\mathbf x$, and $h_j(y)$ its $q$-weighted average over voxel $j$. Thus $\int_{\mathcal Y}h_j(y)\,dy=1$ on the selected observation space. Writing $a_{ij}=q_jh_j(y_i)$, the Poisson point-process model is
\begin{align}
 m_i&=\sum_j a_{ij}\lambda_j+\sum_c B_c h_c(y_i),\label{eq:intensity}\\
 \ell(\bm\lambda,\mathbf B)&=\sum_i\log m_i-\sum_jq_j\lambda_j-\sum_c B_c,
 \label{eq:lik}
\end{align}
where $Y=(y_1,\dots,y_n)$, $m_i$ is the expected event density at $y_i$ summed over all sources, $\lambda_j\ge0$ is voxel activity, $B_c$ is the expected count for background class $c$, and $h_c$ is its normalized density.

The working response factors the true-event energy density $p_E$ and a spatial kernel, $h(y_i\mid\mathbf x)\propto p_E(\E_i)k_i(\mathbf x)$. The Gaussian approximation uses the covariance in \eqref{eq:cov}; the refined kernel is
\begin{equation}
\begin{split}
 k_i(\mathbf x)&=\sum_{l=1}^{L}\pi_l\phi_3\{\Vh_i-\mathbf x; c_l\bS_i\},\\
 &\pi_l\ge0,\quad \sum_l\pi_l=1,\quad c_l>0.
\end{split}
 \label{eq:mixture}
\end{equation}
Here $\phi_3(\cdot;\bS)$ is the trivariate normal density with covariance $\bS$, $\bS_i=\bS_E+\bS_H$ the covariance \eqref{eq:cov} of event $i$, and $\pi_l$, $c_l$ the weights and scale factors; the mixture parameters are fitted to labeled calibration events independently of the evaluated events. This approximate response must be normalized over the selected observations, which is checked by simulation; normalizing over source coordinates alone does not suffice. Sensitivity is estimated after energy smearing and selection (selected fractions 0.423 and 0.402 for a central and an off-axis compartment).

For a known region $R_r$ with constant three-photon activity density $\theta_r\equiv\theta^{3\gamma}_r$ (three-photon decays per unit volume during the acquisition), define $Q_r=\int_{R_r}q(\mathbf x)\,d\mathbf x$ and
\begin{equation}
 F_{ir}=Q_r^{-1}\int_{R_r}q(\mathbf x)h(y_i\mid\mathbf x)\,d\mathbf x,
 \qquad \nu_r=Q_r\theta_r.\label{eq:region}
\end{equation}
The regional intensity is $m_i=\sum_r\nu_rF_{ir}+\sum_c B_ch_c(y_i)$ and its integrated rate is $\sum_r\nu_r+\sum_cB_c$. Known phantom boundaries provide structural information.

Independent Gamma shape--rate priors yield conjugate updates after introducing an event-origin label $z_i$. For the voxel model,
\begin{equation}
\begin{split}
 \Pr(z_i=j\mid y_i,\bm\lambda,\mathbf B)&=\frac{a_{ij}\lambda_j}{m_i},\\
 \lambda_j\mid\mathbf z&\sim\mathrm{Gamma}(\alpha_j+n_j,\beta_j+q_j).
\end{split}
 \label{eq:assign}
\end{equation}
Here $n_j$ is the number of events allocated to voxel $j$ ($n_r$, $n_c$ likewise for regions and background classes) and $\alpha_j,\beta_j$ are the prior shape and rate. Background counts have analogous updates with rate $b_c+1$. Regional expected counts use $\nu_r\mid\mathbf z\sim\mathrm{Gamma}(a_r+n_r,b_r+1)$. The regional analyses use $a_r=1/2$, $b_r=0$, a half-count power prior. It is the Jeffreys form for separately observed Poisson counts, not for overlapping mixture components. Conditional updates and their normalization are given in Supplementary Section~\ref{S-app:gibbs}.

\emph{Hierarchy and posterior.} With rates $\mathbf u=(\bm\nu,\mathbf B)$, responses $H_{ik}\in\{F_{ir},h_c(y_i)\}$, integrated responses $q_k$ (one for normalized components) and hyperparameters $\bm\xi$,
\begin{equation}
p(\mathbf z,\mathbf u,\bm\xi\mid Y)\propto\textstyle\prod_iH_{iz_i}\prod_ku_k^{n_k}e^{-q_ku_k}\,p(\mathbf u\mid\bm\xi)\,p(\bm\xi):
\label{eq:post}
\end{equation}
the measurement model given each event's origin, the Poisson thinning $n_k\mid u_k\sim\mathrm{Poisson}(q_ku_k)$, the conditional prior and the hyperprior. Summing over $\mathbf z$ returns \eqref{eq:lik}; the meaning of each factor and the sampler are detailed in Supplementary Section~\ref{S-app:gibbs}.

\emph{Alternative priors.} Four alternatives keep the likelihood and the Gibbs allocation step. (i) A global--local prior for voxel intensities, $\lambda_j\mid b_j\sim\mathrm{Gamma}(a,b_j)$, $b_j\mid\phi\sim\mathrm{Gamma}(c,\phi)$, with a half-Cauchy-type global scale $\phi$ (the TPBN family \cite{armagan2011}; $a=c=1/2$ is the horseshoe analog \cite{carvalho2010}), so that all conditionals remain Gamma. Such polynomial-tailed global--local priors attain posterior contraction in sparse exponential-family models, including count responses \cite{wang2025mtmbsp}. (ii) A Gaussian Markov random field on the log intensities with a half-Cauchy smoothing scale. (iii) For the region model, shrinkage of the rods' $\log\eta$ toward a common value, $\log\nu_r=\log(Q_r\theta^{2\gamma}_r)+\mu+\delta_r$, where $\theta^{2\gamma}_r$ is the fitted two-photon activity density, $\mu$ a common level (flat prior) and $\delta_r\mid\tau,\psi_r\sim N(0,\tau^2\psi_r)$, so that $\eta_r=e^{\delta_r-\delta_0}$; the global scale is $\tau\sim C^+(0,1)$ and the local scales are $\psi_r=1$ (Gaussian) or $\psi_r^{1/2}\sim C^+(0,1)$ (horseshoe), updated through inverse-gamma auxiliaries \cite{makalic2016}. (iv) A physics-informed version: through the three-photon fraction \cite{jasinska2016}, the o-Ps contrast $\delta=\log(I\tau_{\mathrm{oPs}})-\log(I\tau_{\mathrm{oPs}})_{\mathrm{water}}$ (intensity $I$, lifetime $\tau_{\mathrm{oPs}}$) gives $\eta_r=g(\delta_r)/g(\delta_0)$, $g(\delta)=w_0+(1-w_0)e^{\delta}$, with $w_0\approx0.4$ the direct-annihilation share of three-photon decays in water (a physical floor); $\delta_r$ is shrunk toward water with $\tau\sim C^+(0,0.3)$, a scale set by tissue contrasts \cite{moskal2023myxoma}. Details are in Supplementary Section~\ref{S-sec:s-priors}.

The enhancement ratio relative to reference region $0$ is
\begin{equation}
 \eta_r=\frac{\theta_r^{3\gamma}/\theta_r^{2\gamma}}
 {\theta_0^{3\gamma}/\theta_0^{2\gamma}}.\label{eq:eta}
\end{equation}
Here $\theta^{3\gamma}_r$ and $\theta^{2\gamma}_r$ are the three- and two-photon activity densities of region $r$, and region 0 is the reference (water in the porous phantom, the 5~mm rod in the four-rod phantom). The ratio $\theta^{3\gamma}/\theta^{2\gamma}=f_3/(1-f_3)$ follows the three-photon fraction, and the normalization cancels absolute activity and detection efficiency. We evaluate this functional for every posterior draw and report the posterior median and equal-tailed 95\% interval. The image summary is a posterior mean. In simulations, two-photon uncertainty is approximated on the log-intensity scale; in the measured analysis, fitted two-photon intensities are fixed. Reported experimental intervals are therefore conditional statistical intervals.

\subsection{Maximum Likelihood, MLEM, and Loss}\label{sec:theory}
For fixed response and background, the same likelihood yields the MLEM update
\begin{equation}
 \lambda_j^{(t+1)}=\frac{\lambda_j^{(t)}}{q_j}
 \sum_i\frac{a_{ij}}{m_i^{(t)}}.\label{eq:mlem}
\end{equation}
Regional MLE uses the class densities and region geometry of the Bayesian fit with expected instead of sampled allocations; it differs from voxel MLEM stopped at 50 iterations, whose iteration number acts as an implicit regularizer.

\emph{ML as a limit of the posterior.} EM is a minorize--maximize algorithm \cite{lange2000} whose M-step $u_k=\sum_iw_{ik}/q_k$, with expected labels $w_{ik}$, is the mode of the Gibbs conditional $\mathrm{Gamma}(a_k+n_k,b_k+q_k)$ under the flat prior $a_k=1$, $b_k=0$: ML3 is the flat-prior posterior mode with sampling replaced by expectation. With a fixed number of parameters, an interior truth and many counts per class, the posterior is approximately normal about the MLE \cite{vandervaart1998}; the median of $\log\eta_r$ then differs from the MLE by about $\tfrac16(1/n_r-1/n_0)$, against a posterior SD of about $(1/n_r+1/n_0)^{1/2}$. This holds for the regional model, not for voxel images, whose dimension is comparable to the counts, nor at the boundary $\lambda_j=0$, where the difference is of the order of the SD (Supplementary Section~\ref{S-app:gibbs}).

For any loss, the Bayes action minimizes the posterior expected loss and, under a proper prior, has the smallest Bayes risk; it need not have lower risk at every truth. Squared error is not natural for Poisson intensities, whose variance grows with the mean, or for ratios. Poisson-adapted losses give other Bayes actions (Supplementary Table~\ref{S-tab:s-loss}): for a count $N\sim\mathrm{Poisson}(s\theta)$ with known exposure $s$, the Kullback--Leibler loss
\begin{equation}
 L_{\mathrm{KL}}(\theta,d)=s\{d-\theta+\theta\log(\theta/d)\}=\mathrm{KL}\{\mathrm{Po}(s\theta)\,\|\,\mathrm{Po}(sd)\}
 \label{eq:klloss}
\end{equation}
gives the posterior mean, the squared log error the geometric mean and Stein's loss $1/\mathbb E[\theta^{-1}\mid Y]$. For $\eta$ we report the posterior median, the Bayes action for the absolute log error; it requires only $\mathbb E|\log\eta|<\infty$, whereas the posterior mean of a ratio may not exist.

The loss decides whether a prior can dominate the MLE (Supplementary Section~\ref{S-sec:s-risk}). For $N\sim\mathrm{Poisson}(s\theta)$ the half-count posterior mean $(N+1/2)/s$ has a mean squared error larger than that of $N/s$ by $1/(4s^2)$, yet under \eqref{eq:klloss} the MLE, zero with probability $e^{-s\theta}$, has infinite risk and every Gamma posterior mean with positive shape dominates it. For $p\ge2$ Poisson means, shrinkage dominates the MLE under normalized squared error \cite{clevenson1975} (risk \CZLow\ against 4 at means $(0.2,0.5,1,2)$, \CZHigh\ at $(50,\dots,80)$), so gains are appreciable only at a few expected counts per parameter (voxels, not regions).

\begin{figure*}[!t]
\centering\includegraphics[width=\textwidth]{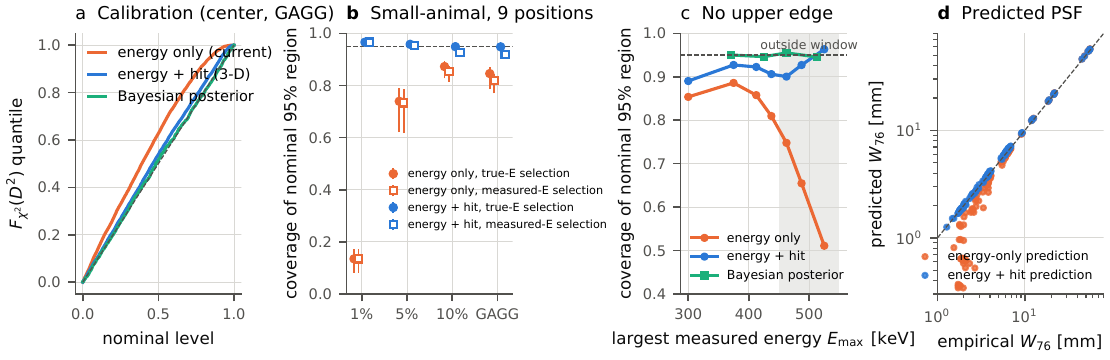}
\caption{Event-level calibration. (a) Quantiles of $F_{\chi^2}(D^2)$ against the nominal level (small-animal center, GAGG, measured-energy selection). (b) Coverage of the nominal 95\% region for the energy-only and energy+hit covariance (mean and range over nine positions; filled: true-energy, open: measured-energy selection). (c) Coverage by measured $E_\text{max}$ without an upper edge (50--550~keV). (d) Predicted vs.\ empirical $W_{76}$.}
\label{fig:cal}
\end{figure*}
\subsection{Simulation Design}\label{sec:sim}\label{sec:g4like}
Two simulation levels separate ideal response calibration from transport and background effects. The full-absorption surrogate generates Ore--Powell photon energies and isotropic decay planes \cite{orepowell1949}, models pixel geometry and photon interactions, and applies measured-energy noise and selection. The HR-GAGG modules contain $8\times8$ pixels of $2.5\times2.5\times9$~mm$^3$ at 3.2~mm pitch. The measured energy response is FWHM$/E=513.3 E^{-1.94}+0.104$, with $E$ in keV. Scanner geometries, idealized response variants, and validation against the companion scanner study \cite{fujimoto2026bams} are specified in Supplementary Section~\ref{S-sec:s-mc}. Reported radial, tangential, and axial widths agree within \ValDev\%.

A porous-material phantom has water and XAD-4 compartments at $(\mp15,0)$~mm with $\eta=1.78$, the value expected from positronium lifetimes in XAD-4 \cite{fujimoto2026commphys}, 300, 1000, or 3000 expected selected signal events, with and without two-photon scatter, and a configuration with water at the center; a four-compartment simulation evaluates slice-wise and volumetric reconstructions (Supplementary Table~\ref{S-tab:p3d}).

The transport-based four-rod simulator, a custom Monte Carlo model (not Geant4), includes $^{18}$F positron range, water and PLA walls, partial deposition, inter-pixel and inter-module Compton scatter, Gd fluorescence, energy smearing, timing walk, and accidental coincidences. The phantom comprises 2, 3, 4, and 5~mm diameter water cavities with 2~mm walls and bottoms on a 3~mm base; the cavities are 7~mm deep. The simulator sums deposits within modules, whereas the measurement sorts pixel hits.

Pools of $2.5\times10^7$ three-photon and $5\times10^7$ two-photon decays per rod are simulated once; each dataset is a Poisson resample of the pools with independent read-out noise, so pool fluctuations are common to all datasets of a setting; five settings were repeated with six independent pool realizations (Supplementary Table~\ref{S-tab:s-pools}). Labeled templates from a separate pool calibrate the response and background. Random triples are scaled to a late-to-prompt ratio of \GLLatePrompt\ for measured triples from three distinct modules, a working assumption. Thirty datasets per setting evaluate background, count level, contrast, wall fraction, response mismatch, and template size. Full generation and selection details are in Supplementary Section~\ref{S-sec:s-g4like}.

\subsection{Measured and Geant4 Four-Rod Data}\label{sec:realdata}
Fujimoto \emph{et al.} measured four rods of 2, 3, 4 and 5~mm diameter at 15~mm from the axis, filled with $^{18}$F-FDG of uniform concentration (2.45~MBq in total), with a two-ring prototype (16 modules, 1024 pixels) \citep{fujimoto2026js}.
Because every rod holds the same solution, $\eta=1$ for every rod unless annihilations in the rod walls change the three-photon fraction. The 3D-printed cavities are about 7~mm long; effective lengths fitted from the two-photon data are model parameters.
The data comprise \NTriplesExp\ triple coincidences (100--450~keV per photon, total 920--1124~keV, $t_{13}<100$~ns), \NPairsExp\ pairs (485--537~keV) and \NSimTriples\ Geant4 triples with the decay mode of each photon (\NSimTrue\ three-photon). Unlike the phantom, the Geant4 model has polyvinyl alcohol walls, 250~keV positrons, ortho-positronium only in water (two-photon wall annihilations) and no base plate (Supplementary Section~\ref{S-sec:s-real}).
The triples were sorted from pixel hits without module summation, with one time offset per module and no walk correction; two hits in one module occur in \RealSameModAll\% of the prompt triples with $E_\text{tot}\ge985$~keV before the vertex selection and \RealSameModBox\% after it. Pixel numbering and module orientations were inferred from the two-photon likelihood and confirmed by the detector group (Supplementary Section~\ref{S-sec:s-real}).

\emph{Two-photon model.} Pairs are binned by pixel pair (\NUniqueLORs\ distinct pairs) in a prompt ($|\Delta t|<20$~ns) and a delayed (50--100~ns) window. The line-of-response density for a decay in rod $r$ is a Gaussian tube (SD 1.2~mm) integrated over the rod, and a diffuse cylinder absorbs activity outside the rods. EM gives the rod counts, with randoms removed through the delayed window; rod centers and axial extents were fitted by profile likelihood.

\emph{Three-photon model.} For prompt triples ($t_{13}<40$~ns, $E_\text{tot}\ge985$~keV, valid vertex in a $56\times56\times28$~mm$^3$ box) we use the region model with the 3-D kernel $\mathcal{N}(\Vh_i,\bS_E+\bS_H)$, a selection-aware axial sensitivity from the surrogate, and three non-rod classes with normalized energy--position densities: a uniform \emph{diffuse three-photon} class that absorbs events reconstructed far from every rod (about 30\% of the true three-photon events in Geant4), a \emph{two-photon scatter} class from the Geant4 events labeled two-photon, and a \emph{randoms} class from the late window $60\le t_{13}<100$~ns, treated as random-dominated (for three different modules, $t_{13}$ is flat within \RealLateFlat\% beyond 40~ns). Temporal flatness alone does not exclude delayed true events, so late-window contamination is an unresolved source of systematic uncertainty.
The energy density $p_E$ is a kernel estimate on the Geant4 three-photon events. $\bS_E$ and $\bS_H$ were scaled by factors chosen by the likelihood on the measured events (1.5 and 2.0), which also absorb PSF tails and residual geometry errors.

\emph{Refinements from the simulation.} The simulator showed that the Gaussian kernel misses heavy-tailed vertex errors (Sec.~\ref{sec:res-g4like}); the refined model B3 used in the simulation study (1) replaces the kernel by a scale mixture $\sum_k\pi_k\mathcal{N}(\Vh_i,c_k(\bS_E+\bS_H))$ fitted by EM to the true vertex errors of labeled simulations, without a diffuse class; (2) anchors the scatter class to the rods, using the template distribution of the reconstructed vertex about the decay rod weighted by the two-photon counts; (3) blurs the rods by the $^{18}$F positron range; and (4) enters the expected randoms of each pixel pair as a known term in the two-photon EM, with pair efficiencies from the simulation (Supplementary Section~\ref{S-sec:s-g4like}).
B3-G, without refinement 1 (Gaussian kernel and diffuse class), is the study's counterpart of the initial measured-data model; the heavy-tailed measured-data variants introduce refinement 1 only.
A complete three-photon fit takes about 12~s on two CPU cores \citep{lam2015numba} (Supplementary Section~\ref{S-sec:s-fast}).
The ROI estimator (M1) is our implementation of the method of \citep{fujimoto2026commphys}: kernel deposition with $\bS_E$ for $E_\text{tot}\ge985$~keV and $t_{13}<100$~ns, a two-photon MLEM image blurred to the three-photon resolution, ROIs of radius $r_\text{phys}+f\times\text{FWHM}$ ($f=0.5$, 1; FWHM 4.1~mm from the median covariance), and $\eta$ from the ROI count ratios.

\subsection{Estimators and Evaluation}\label{sec:eval}
\emph{Event level}: (a) energy term with exact hits; (b) hit term with true energies; (c) current practice, pixel hits with the in-plane $\bS_E$; (d) $\bS_E+\bS_H$; plug-in Monte Carlo propagation; and the event posterior. Coverage is the fraction of events with squared Mahalanobis distance below the $\chi^2_d$ quantile; it checks each event's Gaussian kernel, not the Poisson counts (Supplementary Section~\ref{S-sec:s-cov}).
\emph{ROI level}: M1, our implementation of the published-style pipeline (985~keV cut, $\bS_E$ deposition, sensitivity-normalized two-photon image, ROI radius $2\sigma$ of each region's width, Poisson interval); M2, list-mode MLEM with $\bS_E+\bS_H$ and sensitivity; B2, the voxel model with background class; B3, the region model (with background class, all selected events); and, for the four-rod phantom, BS, an emulation of a system-matrix method with an energy-neighborhood filter and independent 2-D reconstructions per $z$ slice. ML3 is the regional MLE under the same event-by-class likelihood matrix, sensitivities and two-photon denominator as B3 in the same dataset: EM with SQUAREM \cite{varadhan2008squarem} to a relative change below $10^{-9}$, with a delta-method 95\% interval on $\log\eta$ from the observed information (Supplementary Section~\ref{S-app:gibbs}). B3-mean is the posterior mean of $\eta$. On the measured data a cluster bootstrap resamples the groups of triples that share a hit.
\emph{Simulated four-rod study} (Sec.~\ref{sec:g4like}): for each setting of Table~\ref{tab:g4like}, 30 datasets of $T=4000$~s, which hold about \GLNThreeStudy\ selected true three-photon triples (about a quarter as many at $T/4$ and with the module response). Each dataset has its own labeled template simulation, so the template noise enters the comparison. The methods are B3, B3-G, B3 without the scatter and randoms classes (rods and a diffuse class only), ML3 and B3-mean under the likelihood of B3, and M1 with ROIs of $r+0.5$ and $r+1$~FWHM.

\section{Results}

\subsection{Event-Level Coverage and the Hit-Position Term}\label{sec:res-event}
\begin{table*}[t]\centering\small
\caption{Event-level coverage (nominal 68\%/95\%) of the propagated covariance, measured GAGG response, averaged over source positions ($n_\text{pos}$). (a) energy term with exact hits; (b) hit term with true energies; (c) current practice: pixel hits, energy-only $2\times2$ in-plane covariance; (d) proposed: pixel hits, $\Sigma_E+\Sigma_H$ (3-D). Axial: coverage of the 1-D 95\% interval on $z$. Last column: $W_{76}$ (radial) predicted from the covariances of (d) divided by the empirical $W_{76}$.}\label{tab:event}
\resizebox{\textwidth}{!}{\begin{tabular}{llrcccccc c}\toprule scanner & selection & $n_\text{pos}$ & (a) $E$, exact hits & (b) hits, true $E$ & (c) current & (d) $E$+hit & axial (c) & axial (d) & pred./emp.\\\midrule
small & true $E$ & 9 & 0.68/0.92 & 0.66/0.97 & 0.55/0.85 & 0.69/0.95 & 0.89 & 0.96 & 1.02 \\
small & measured $E$ & 9 & 0.64/0.90 & 0.66/0.97 & 0.52/0.82 & 0.65/0.92 & 0.86 & 0.94 & 0.95 \\
small & no upper edge & 9 & 0.63/0.86 & 0.65/0.96 & 0.45/0.72 & 0.67/0.92 & 0.88 & 0.94 & 1.04 \\
brain & true $E$ & 3 & 0.68/0.92 & 0.66/0.97 & 0.66/0.92 & 0.67/0.93 & 0.95 & 0.96 & 1.01 \\
brain & measured $E$ & 3 & 0.64/0.89 & 0.66/0.97 & 0.62/0.89 & 0.63/0.90 & 0.92 & 0.93 & 0.93 \\
human & true $E$ & 3 & 0.68/0.92 & 0.66/0.97 & 0.68/0.92 & 0.67/0.93 & 0.95 & 0.96 & 1.03 \\
human & measured $E$ & 3 & 0.64/0.89 & 0.65/0.97 & 0.64/0.89 & 0.63/0.90 & 0.92 & 0.93 & 0.93 \\
\bottomrule\end{tabular}}\end{table*}

With exact hits, the energy-only covariance covers \CovAexactSixtyEight\ of events at the nominal 68\% and \CovAexact\ at 95\% (Table~\ref{tab:event}, Fig.~\ref{fig:cal}).
With pixel hits it covers only \CovCurSmallTrueNF\ in the small-animal scanner, and the deficit grows as the energy resolution improves: \CovCurOnePct\ at 1\% and \CovCurFivePct\ at 5\%.
At the GAGG response the hit term accounts for \HitShareGAGGxy\ of the in-plane and \HitShareGAGGz\ of the axial variance.
Adding $\bS_H$ restores the 95\% coverage to \CovEHrange\ for all energy models and positions, and the axial 95\% interval to \AxialEH\ (energy-only \AxialE).
The PSF predicted by drawing one error per event from $\mathcal{N}(0,\bS_i)$ reproduces the empirical $W_{76}$ with a median ratio of \PredEmpEH\ (energy-only: \PredEmpE), a test applicable to measured point sources.
Selecting on measured energies lowers the 95\% coverage to \CovEHmeas\ (\CovEHwide\ without an upper edge), because events with true $E_\text{max}>450$~keV enter with better-conditioned measured energies. The first-order covariance agrees with Monte Carlo propagation (SD ratio 1.02--1.04), an auxiliary event posterior with Gaussian energy errors and latent hit positions (Supplementary Section~\ref{S-sec:s-bayes}) covers 0.947--0.957, and a common gain error leaves every vertex unchanged (Supplementary Tables~\ref{S-tab:jmb} and~\ref{S-tab:energy}).

\subsection{Kernel Deposition and Voxel-Level Uncertainty}\label{sec:res-image}
Depositing each event with its energy-only kernel broadens the image by $\sqrt{1+f_E}$, $f_E$ the energy share of the variance, rather than by $\sqrt2$, because the hit term is missing from the kernel (Supplementary Section~\ref{S-app:sqrt2}). This predicts 1.386 for GAGG against 1.375 in the companion Geant4 study \citep{fujimoto2026bams} and \DepOurs\ here, with no free parameter (Fig.~\ref{S-fig:sqrt2}).
In the voxel model the 95\% intervals of single 2~mm voxels cover the truth in only 32\% of cases (28\% with background), so voxelwise uncertainty is not calibrated at this resolution, whereas the activity in a 6~mm disc is recovered to 99\% (97\%) (Fig.~\ref{S-fig:images}).

\begin{figure*}[!t]
\centering\includegraphics[width=0.9\textwidth]{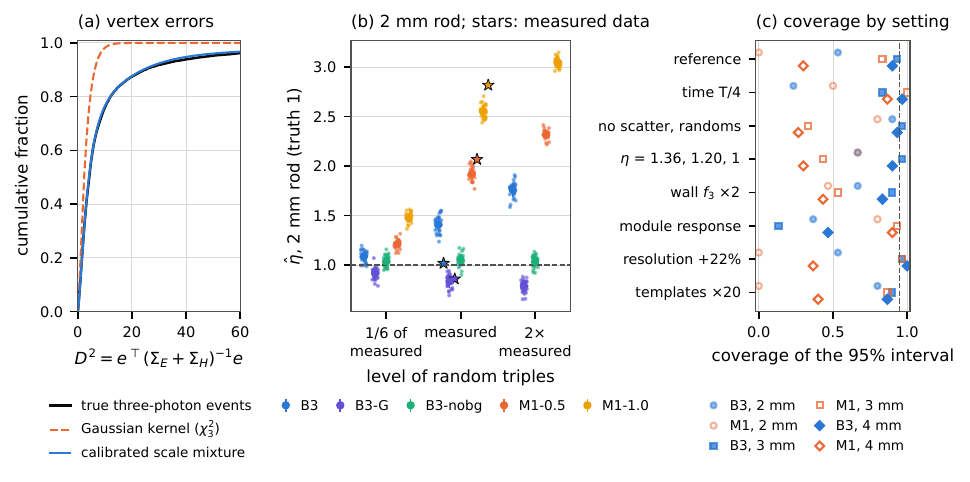}
\caption{Transport-based simulation of the four-rod measurement. (a) Squared Mahalanobis distance of the true vertex errors of selected three-photon triples under $\bS_E+\bS_H$, against the Gaussian ($\chi^2_3$) and the calibrated scale-mixture predictions. (b) Estimates of $\eta$ for the 2~mm rod (truth 1) over 30 datasets of one pool realization at three levels of random triples (points; mean $\pm$ SD), with the measured values of the closest counterparts (stars; B3: heavy-tailed kernel without refinements 2--4; B3-G: measured-data model). (c) Coverage of the 95\% intervals of B3 (filled) and M1 at $r+0.5\,$FWHM (open) for the 2, 3 and 4~mm rods, by setting with randoms at 1/6 of the measured level or none.}
\label{fig:g4like}
\end{figure*}
\subsection{Porous-Material Phantom}\label{sec:res-roi}
\begin{table}[!t]\centering\footnotesize\setlength{\tabcolsep}{3pt}
\caption{Porous-material simulation ($\eta=1.78$): repeated-dataset results, 200 replicates per displayed configuration. M2 is 50-iteration voxel MLEM; ML3 is matched regional MLE; B3 is Bayesian regional inference. Values are on the ratio scale.}\label{tab:eta}
\begin{tabular}{rlrrrr}\toprule
Events & Method & Bias & SD & RMSE & Cov. 95\%\\\midrule
3000 & M1 & +0.015 & 0.074 & 0.075 & 0.97\\
& M2 & $-$0.016 & 0.093 & 0.094 & --\\
& ML3 & $-$0.010 & 0.066 & 0.067 & 0.96\\
& B3 & $-$0.010 & 0.067 & 0.067 & 0.96\\\midrule
1000 & M1 & +0.022 & 0.135 & 0.136 & 0.96\\
& M2 & +0.151 & 0.202 & 0.252 & --\\
& ML3 & +0.005 & 0.117 & 0.117 & 0.95\\
& B3 & +0.005 & 0.117 & 0.117 & 0.95\\\midrule
300 & M1 & +0.012 & 0.258 & 0.258 & 0.96\\
& M2 & $-$0.055 & 0.335 & 0.338 & --\\
& ML3 & +0.010 & 0.225 & 0.225 & 0.96\\
& B3 & +0.009 & 0.225 & 0.224 & 0.96\\
\bottomrule\end{tabular}
\end{table}

Table~\ref{tab:eta} compares the estimators of $\eta$ over repeated datasets (all estimators and figures: Supplementary Section~\ref{S-sec:s-extra}).
In the symmetric phantom without background and with 3000 events every estimator is within 1.5\% of $\eta_\text{true}$. The published-style ROI pipeline (M1) has bias +0.8\% and a slightly conservative interval (coverage 0.97), because it treats the ROI integrals of a kernel-deposited image as independent Poisson counts.
The region model (B3), which uses all selected events with their likelihood weights, has bias $-0.5$\%, an SD about 10\% smaller than M1 and 95\% coverage of 0.95--0.96 from 300 to 3000 events, with intervals 18--19\% narrower.
The 3000-event RMSE is 0.067 for both B3 and matched regional MLE (ML3), versus 0.094 for 50-iteration voxel MLEM (M2). M2 also differs in events, grid and ROIs; with those of the voxel Bayesian model, 50-iteration MLEM matches B3 (Sec.~\ref{sec:res-priors}).
With the water compartment at the center, efficiency and recovery no longer cancel: M1 is biased by $-3.3$\% and covers 0.87, whereas B3, which estimates activity densities with the selection-aware sensitivity, has bias +0.0\% and coverage 0.93.
With one background event per true event, the background class reduces the bias from $-2.8$\% to $-0.6$\% (coverage 0.86 to 0.94).

\subsection{Transport-Based Simulation of the Four-Rod Measurement}\label{sec:res-g4like}\label{sec:res-3d}
\begin{table*}[!t]\centering\footnotesize\setlength{\tabcolsep}{3.2pt}
\caption{transport-based simulation of the four-rod measurement: relative bias [\%] of $\hat\eta$ and coverage of the 95\% interval for the 2, 3 and 4~mm rods (relative to the 5~mm rod) over repeated datasets (number in parentheses) of one pool realization (six realizations: Table~\ref{S-tab:s-pools}). Each dataset has $T=4000$~s at 2.45~MBq (about 9\,000 selected true three-photon triples; four times fewer at $T/4$ and with the module response) and its own Geant4-sized labeled template simulation. Reference: $\eta=1$, random triples at 1/6 of the measured level. B3: region model with the heavy-tailed kernel calibrated on the template simulation; B3-G: Gaussian kernel and diffuse class (closest counterpart of the measured-data model); M1: ROI estimator. All methods and statistics: Table~\ref{S-tab:s-g4full}.}\label{tab:g4like}
\begin{tabular}{lrrrcccrrrrrr}\toprule
 & \multicolumn{3}{c}{B3 bias [\%]} & \multicolumn{3}{c}{B3 coverage} & \multicolumn{3}{c}{B3-G bias [\%]} & \multicolumn{3}{c}{M1 ($r+0.5\,$FWHM) bias [\%]}\\
\cmidrule(lr){2-4}\cmidrule(lr){5-7}\cmidrule(lr){8-10}\cmidrule(lr){11-13}
setting (datasets) & 2 & 3 & 4~mm & 2 & 3 & 4~mm & 2 & 3 & 4~mm & 2 & 3 & 4~mm\\\midrule
reference (30) & +8.8 & +0.7 & $-$1.0 & 0.53 & 0.93 & 0.90 & $-$8.3 & $-$5.6 & $-$3.0 & +21 & $-$5 & $-$8\\
acquisition time / 4 (30) & +28.2 & +6.9 & +1.4 & 0.23 & 0.83 & 0.97 & $-$3.7 & $-$3.5 & $-$1.0 & +22 & $-$4 & $-$7\\
no scatter or randoms (30) & +1.9 & +1.4 & $-$1.0 & 0.90 & 0.97 & 0.93 & $-$3.9 & $-$2.0 & $-$2.4 & $-$6 & $-$10 & $-$8\\
$\eta=1.36, 1.20, 1$ (30) & +6.1 & +0.5 & $-$1.3 & 0.67 & 0.97 & 0.90 & $-$7.0 & $-$3.6 & $-$3.0 & +8 & $-$8 & $-$7\\
wall three-photon fraction $\times2$ (30) & +6.0 & +0.3 & $-$0.8 & 0.67 & 0.90 & 0.83 & $-$8.2 & $-$5.0 & $-$2.6 & +12 & $-$7 & $-$6\\
module response 0.3--1 (30) & +24.5 & +19.4 & +12.5 & 0.37 & 0.13 & 0.47 & $-$12.4 & +6.5 & +8.9 & +15 & +6 & +4\\
resolution 22\% wider than assumed (30) & +10.2 & +1.2 & $-$0.8 & 0.53 & 0.97 & 1.00 & $-$11.1 & $-$5.7 & $-$3.6 & +33 & $-$2 & $-$7\\
templates $\times20$ (30) & +7.3 & +1.0 & $-$1.0 & 0.80 & 0.90 & 0.87 & $-$7.8 & $-$4.1 & $-$2.9 & +21 & $-$4 & $-$7\\
randoms at the measured level (30) & +41.1 & +7.6 & +2.5 & 0.00 & 0.33 & 0.87 & $-$16.3 & $-$10.8 & $-$3.8 & +92 & +16 & $-$3\\
randoms $2\times$ measured (30) & +76.0 & +18.3 & +6.0 & 0.00 & 0.00 & 0.37 & $-$21.3 & $-$11.1 & $-$5.3 & +132 & +31 & $-$2\\
\bottomrule\end{tabular}\end{table*}

\emph{Vertex errors.} The propagated covariance does not describe the vertex errors of realistic data. Under $\bS_E+\bS_H$ the selected true three-photon triples are covered at 68\% and 95\% in only \GLGaussCovSixtyEight\ and \GLGaussCov\ of cases (Fig.~\ref{fig:g4like}a); their in-plane errors have a median of \GLErrMed~mm but 90th and 99th percentiles of \GLErrNinety\ and \GLErrNinetyNine~mm (scatter in the detector and the phantom). A three-component scale mixture fitted on a 20 times larger labeled simulation ($\pi$ = \GLMixPi; $c$ = \GLMixC) covers \GLMixCovSixtyEight\ and \GLMixCov\ on independent events; in the study B3 refits it on each dataset's template.

\emph{Moderate randoms.} Table~\ref{tab:g4like} summarizes the study.
\begin{itemize}
\item \emph{Reference setting.} B3 has biases of \GLReferenceBBias\% for the 2, 3 and 4~mm rods and 95\% coverage of \GLReferenceBCov; without background triples, \GLNoBackgroundBBias\% (\GLNoBackgroundBCov). Six-realization means: \PRRefBBias\% (\PRRefBCov) and \PRNoBgBBias\% (\PRNoBgBCov). Each background class alone biases the 2~mm rod (scatter only: \GLDiagRefNoRandTwo\%; randoms only: \GLDiagRefNoScatTwo\%).
\item \emph{B3-G} is biased by \GLReferenceGBias\%: the Gaussian kernel and diffuse class move tail events out of the rods.
\item \emph{The ROI estimator} is biased by \GLReferenceMhBias\% ($r+0.5\,$FWHM) and \GLReferenceMfBias\% ($r+1\,$FWHM); its Poisson interval covers \GLReferenceMhCov\ and \GLReferenceMfCov.
\item \emph{Other settings} (Table~\ref{S-tab:s-g4full}) keep the bias pattern of the reference; with templates 20 times larger the 2~mm bias is \GLLargeTemplatesBTwo\% (coverage \GLLargeTemplatesBCov).
\item \emph{Two settings bias B3.} At $T/4$ B3 is biased by \GLLowCountsBBias\%, because the late-window randoms template, taken from the dataset, is four times sparser (randoms only: \GLDiagLowNoScatTwo\%); over six pool realizations the 2~mm bias averaged \PRLowBTwo\% (range \PRLowBTwoRange\%; this one is the largest). An unmodeled module response (\GLModuleResponseBBias\%) also lowers the counts fourfold and breaks the proportionality of three- and two-photon sensitivities (without random triples: \GLDiagModNoRand\%).
\item \emph{Matched ML.} In each dataset ML3 maximizes the likelihood that B3 samples. Over the \GLMLNSettings\ settings the relative biases of ML3 and B3 differ by at most \GLMLMaxBiasDiff\ percentage points (\PRMaxBiasDiff\ over six pool realizations) and their RMSE ratio (B3/ML3) is \GLMLRmseRatio\ (Supplementary Table~\ref{S-tab:s-matched-g4}); at the reference ML3 has biases of \GLMLReferenceBias\% and coverage \GLMLReferenceCov. The background biases above are properties of the likelihood, not of the prior.
\end{itemize}

\emph{Measured randoms.} At the measured late-to-prompt ratio (\GLLatePromptSimMeas; measured \GLLatePrompt) every method is biased for the 2 or the 3~mm rod (Fig.~\ref{fig:g4like}b). For the 2~mm rod M1 gives \GLRandomsMeasuredMhTwo\ and \GLRandomsMeasuredMfTwo\ at the two ROI sizes, and \GLRandomsHighMhTwo\ and \GLRandomsHighMfTwo\ at twice the measured level (measured: 2.07 and 2.82; \RealMOneDm\ with three different modules); B3-G gives \GLRandomsMeasuredGTwo\ (measured 0.86, or \RealMainDmTwo\ with three different modules), B3 \GLRandomsMeasuredBTwo, and B3 without background classes \GLRandomsMeasuredNTwo\ (3~mm rod: \GLRandomsMeasuredNThree).
The prompt window then holds \GLRandPerTrueMeas\ (\GLRandPerTrueHigh) random triples per true triple; the measurement has about \RealRandPerTrue\ (Supplementary Section~\ref{S-sec:s-g4like}).

\subsection{Measured Four-Rod Data}\label{sec:res-real}
\begin{table*}[!t]\centering\footnotesize\setlength{\tabcolsep}{4pt}
\caption{Measured four-rod phantom (uniform FDG concentration, so $\eta=1$ for every rod unless the rod walls change the three-photon fraction): enhancement ratio of the 2, 3 and 4~mm rods relative to the 5~mm rod. ROI values have no interval. Bayesian rows: posterior median and 95\% interval of the three-photon concentration ratio divided by the two-photon concentration ratio of the same geometry (statistical only; the two-photon systematic range is given in the text). All variants: Table~\ref{S-tab:s-realall}.}\label{tab:real}
\begin{tabular}{lccc}\toprule & 2~mm & 3~mm & 4~mm\\\midrule
\multicolumn{4}{l}{\emph{ROI estimator (M1, our implementation of the method of \cite{fujimoto2026commphys})}}\\
M1, ROI $r+0.5\,$FWHM & 2.07 & 1.08 & 1.27\\
M1, ROI $r+1.0\,$FWHM & 2.82 & 1.38 & 1.31\\
\midrule\multicolumn{4}{l}{\emph{Bayesian region model (median [95\% interval])}}\\
Gaussian kernel, diffuse class (main): $E_\text{tot}\ge985$~keV, $t_{13}<40$~ns & 0.86 [0.75, 0.98] & 0.68 [0.62, 0.76] & 0.98 [0.93, 1.05]\\
covariance scales not calibrated (1, 1) & 0.81 [0.70, 0.93] & 0.66 [0.59, 0.73] & 0.97 [0.91, 1.03]\\
all $E_\text{tot}$ (920--1124~keV) & 0.55 [0.45, 0.65] & 0.53 [0.48, 0.59] & 0.85 [0.80, 0.90]\\
3$\gamma$ energy density from the surrogate MC & 0.98 [0.86, 1.12] & 0.73 [0.66, 0.81] & 1.03 [0.96, 1.09]\\
no diffuse 3$\gamma$ class & 0.93 [0.82, 1.05] & 0.71 [0.65, 0.77] & 0.99 [0.93, 1.05]\\
heavy-tailed kernel (simulation-calibrated), no diffuse class & 1.02 [0.92, 1.14] & 0.74 [0.67, 0.80] & 1.02 [0.96, 1.08]\\
heavy-tailed kernel, scales (1.5, 2.0), no diffuse class & 1.08 [0.97, 1.20] & 0.76 [0.70, 0.82] & 1.04 [0.98, 1.09]\\
heavy-tailed kernel with diffuse class & 0.84 [0.73, 0.96] & 0.66 [0.60, 0.74] & 1.00 [0.94, 1.07]\\
main $\times$ module-efficiency proxy & 0.74 [0.64, 0.85] & 0.68 [0.61, 0.75] & 0.93 [0.88, 0.99]\\
\bottomrule\end{tabular}\end{table*}

\RealParagraph

\emph{Heavy-tailed kernel.} With the kernel calibrated on the simulation and no diffuse class, the region model gives \RealMix\ for the 2, 3 and 4~mm rods, and \RealMixDiffTwo\ with a diffuse class (Table~\ref{tab:real}). A diffuse class raises the measured log-likelihood by \RealDiffuseDeltaLL, but it does so also on simulated datasets without diffuse activity (by \GLDiffMeasDLL, moving the 2~mm rod from \GLDiffMeasBaseTwo\ to \GLDiffMeasTwo): it absorbs template misfit rather than indicating diffuse activity.
With three different modules, as in the simulation, the model gives \RealMixDm\ (Table~\ref{S-tab:s-realall}), and M1 rises to \RealMOneDm\ for the 2~mm rod, so the same-module triples do not cause the high ROI values.

\emph{Matched maximum likelihood and shared hits.} Maximizing the same likelihood (ML3) changes $\eta$ by at most \RealMLMaxDiff\ in any measured variant; with the heavy-tailed kernel ML3 gives \RealMLMix, with a 2~mm Wald interval of \RealMLMixTwoCI\ against the credible interval \RealBMixTwoCI\ (Supplementary Table~\ref{S-tab:s-matched-real}). The measured deviations from $\eta=1$, including the 3~mm deficit, are therefore properties of the likelihood (kernel, classes, sensitivities), not of the half-count prior. The \RealNEventsSel\ selected triples form \RealNGroupsSel\ hit-sharing groups. A cluster bootstrap over these groups gives SDs \RealClusterPostRange\ times the posterior SDs (Supplementary Table~\ref{S-tab:s-matched-real-sd}), within bootstrap error: hit reuse barely affects the analysis selection.

\begin{figure*}[!t]
\centering\includegraphics[width=\textwidth]{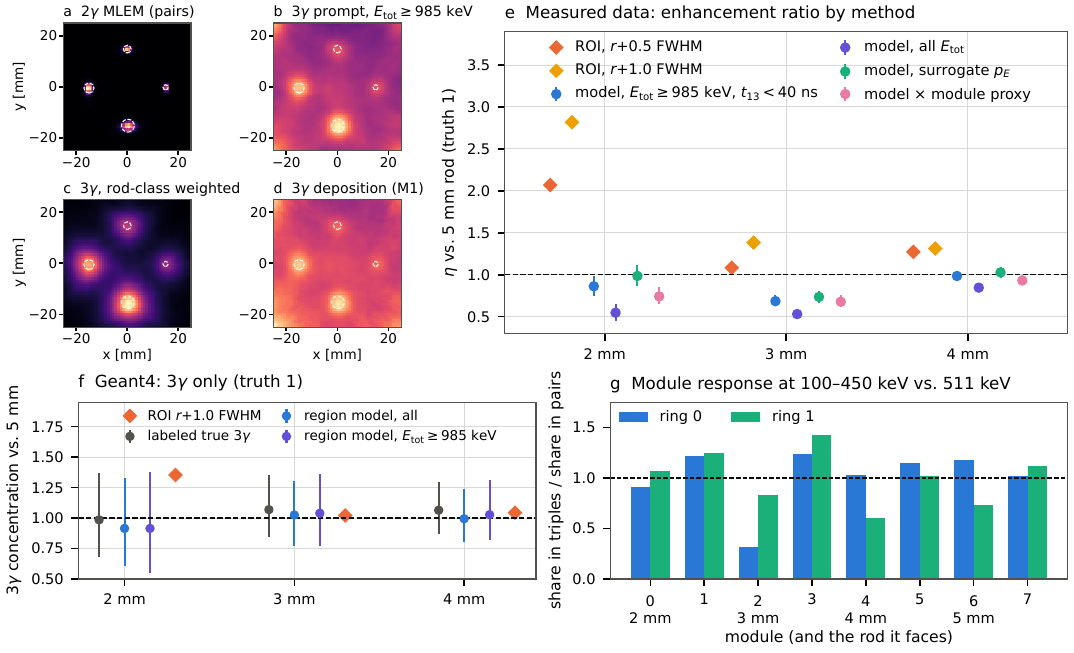}
\caption{\RealCaption}
\label{fig:real}
\end{figure*}

\subsection{Priors and Losses}\label{sec:res-priors}
With identical events, kernel, sensitivity, 2~mm voxels and 6~mm ROIs, $\eta$ in the porous-material phantom was insensitive to the prior: at 1000 events the RMSE was \PrBTwoRmseThou\ (Gamma(0.05)), \PrBTwoHSRmseThou\ (horseshoe-type TPB), \PrBTwoGMRFRmseThou\ (Gaussian MRF), \PrMLTwoRmseThou\ (converged voxel ML) and \PrMLEMRmseThou\ (50-iteration MLEM), within Monte Carlo error also at 300 and 3000 events (Supplementary Table~\ref{S-tab:s-priors-porous}). The images differ: under the Poisson KL error, converged voxel ML is one to two orders of magnitude worse than any posterior mean, as Sec.~\ref{sec:theory} predicts; the Gaussian MRF reduced the error to \PrGMRFvsGammaKLThree--\PrGMRFvsGammaKLThreeThou\ times that of the Gamma(0.05) prior and was best at 300 events (\PrBTwoGMRFKLThree\ vs.\ \PrMLEMKLThree\ for MLEM), whereas 50-iteration MLEM was lower at 1000 and 3000 events and the sparsity-favoring TPB priors were worst (Supplementary Fig.~\ref{S-fig:s-priors}).
In the four-rod study, shrinking the rods' $\log\eta$ toward a common value lowered the RMSE (2, 3, 4~mm) from \FRRefBRmse\ to \FRRefGSRmse\ at the reference and from \FRNoBgBRmse\ to \FRNoBgHSRmse\ without background, where the truth is uniform, and changed it little for strongly unequal ratios (\FRStrBRmse\ vs.\ \FRStrGSRmse; Supplementary Table~\ref{S-tab:s-priors-g4}). The physics-informed horseshoe prior was the most accurate without background (\FRNoBgPHSRmse) but, with its tighter physical scale, cost accuracy for strongly unequal ratios (\FRStrPHSRmse) and in the two-region porous phantom. On the measured data shrinkage moved $\eta$ by at most \RealPriorMaxShift, insensitively to $w_0$.

\section{Discussion}
\emph{Contribution.}
The MLE comparators are statistical benchmarks, not a relabeling of the published pipeline \cite{fujimoto2026commphys}, in which MLEM reconstructs only the two-photon image.

\emph{Source of the improvement.}
Weak-prior regional inference should approach matched ML when information is sufficient, and it does: in every simulated setting and on the measured data B3 and ML3 agree closely, and with matched events, grid and ROIs 50-iteration MLEM matches them as well. B3 and ML3 share the model bias (relative to the Kullback--Leibler projection of the truth onto the model) and the sampling error, and differ by the $O(1/n)$ effect of the half-count prior, below Monte Carlo error down to about \ChkNTwoMin\ counts in the 2~mm rod (Supplementary Sections~\ref{S-sec:s-risk} and \ref{S-app:gibbs}). Gains over M1 and voxel MLEM thus come from the regional likelihood (event weighting, selection-aware sensitivity, known boundaries, background classes) and the event selection, not from the prior; the posterior adds intervals for functionals.

\emph{Where a prior matters.} Regional classes hold hundreds to thousands of events, so no prior acting on each region changes $\eta$; voxels hold few, so the prior governs the image and must match the object. Shrinkage across regions helps when the ratio is shared but does not remove the background bias.

\emph{Measured phantom and calibration toward $\eta=1$.} The heavy-tailed model is closer to one than the ROI analysis for the 2 and 4~mm rods and farther for the 3~mm rod, which faces a module with low relative response and gain drift. On labeled Geant4 data the model recovers the true concentration ratios (Fig.~\ref{fig:real}f), so the deviations arise in the measurement chain or the walls. Calibrations measured independently of the phantom should bring every rod to $\eta=1$: (i) per-pixel energy calibration with gain tracking, walk correction and merging of split hits, so that the module facing the 3~mm rod (response proxy 0.32; Fig.~\ref{fig:real}g) is treated like the others; (ii) module-wise three-photon sensitivities at 100--450~keV from a uniform-cylinder scan, replacing the proxy, which did not remove the deficit (Fig.~\ref{fig:real}e); (iii) kernel widths and tails from point-source scans, instead of the factors fitted on the phantom; (iv) scans at two activities, separating randoms (quadratic in activity) from delayed true triples (linear) in the late window, and with the phantom rotated by $90^\circ$, so that a deficit that follows the module identifies a detector cause. A uniform phantom then serves as the acceptance test: every rod's 95\% interval should contain 1.

\emph{Use in theranostic decisions.} For a decision threshold $c$, such as one for choosing a $\beta$- or an $\alpha$-emitting agent, the posterior gives $\Pr(\eta_r>c\mid\text{data})$ with count, background and response uncertainty. Averaged over six pool realizations, coverage was \PRNoBgBCov\ without background, \PRRefBCov\ at the reference and \PRRandBCov\ at the measured randoms level, where the model itself is biased by \PRRandBTwo\% for the 2~mm rod (range \PRRandBTwoRange\%): neither sampling nor more MLEM iterations repair a misspecified response. Such decisions need lesions well above the resolution and a validated background model.

\emph{Uncertainty and generalizability.}
Measured intervals condition on the fitted geometry, response, templates and two-photon fit.

The simulator omits electron transport, Rayleigh scattering, pile-up, and dead time; template and kernel uncertainty are not fully propagated, thirty replicates give a coverage standard error of about 0.040, and the pool realization adds an error common to all datasets of a setting that the intervals omit (bias SD up to \PRLowBTwoBsd\ percentage points; Supplementary Table~\ref{S-tab:s-pools}). The undercoverage for 2~mm voxels prevents claims of calibrated voxelwise uncertainty, and regional results rely on known phantom structure.

\section{Conclusion}
A Bayesian response-based reconstruction provides regional intensity and enhancement-ratio inference for three-photon annihilation imaging. The regional estimator improves on the ROI and finite-iteration voxel MLEM pipelines; matched regional MLE performs similarly, so the gains derive from the regional likelihood. Priors changed voxel images substantially but regional ratios only when shrunk toward a common value. Because it needs only positron emission, the method applies to the imaging agents of radiotheranostic pairs, where threshold decisions require calibrated intervals. The 3~mm deficit and background sensitivity remain open.

\section*{Funding and Acknowledgments}
Supported in part by National Science Foundation (NSF) grants DMS-1924792 and DMS-2318925 and by the National Institutes of Health (NIH), National Institute of Allergy and Infectious Diseases (NIAID), R01 award R01AI206603. This work was partially supported by JST CREST, Japan, Grant Number JPMJCR26C4.
Analysis code accompanies this work.

\emph{Conflict of interest:} None declared. \emph{AI use:} Claude (Anthropic) assisted with code and editing, checked by the authors.

\bibliographystyle{IEEEtran}
\bibliography{references}
\end{document}